\documentclass[a4paper,10pt]{article}
\usepackage[xdvi]{graphics}
\usepackage{amssymb}
\usepackage{latexsym}
\usepackage{amsmath}
\usepackage{epsf}
\newcommand{\red}{\rho'_{\mbox{\tiny{eff}}}}
\newcommand{\re}{\rho_{\mbox{\tiny{eff}}}}
\def\eqalign#1{\,\vcenter{\openup1\jot\ialign
    {\strut\hfil$\displaystyle{##}$&$\displaystyle{{}##}$
      \hfil\crcr#1\crcr}}\,}
\numberwithin{equation}{section}
\title{A Maximum Mass-to-Size Ratio 
in Scalar-Tensor Theories of Gravity}
\author{
Tooru TSUCHIDA  
\thanks{Electronic address:tsuchida@astro2.sc.niigata-u.ac.jp} \, , 
Go KAWAMURA
\thanks{Electronic address:kawamura@astro2.sc.niigata-u.ac.jp}\,   
and Kazuya WATANABE
\thanks{Electronic address:kazuya@astro2.sc.niigata-u.ac.jp}\\ 
{\em Department of Physics,~Niigata University, Niigata 950-21, Japan.}}
\date{}
\begin{document}
\maketitle


\begin{abstract}
We derive a modified Buchdahl inequality for
scalar-tensor  theories of gravity. In general relativity, Buchdahl
has shown that the maximum value of the mass-to-size ratio, $2M/R$,
is 8/9 for static and spherically symmetric stars 
under some physically reasonable assumptions. 
We formally apply Buchdahl's method to
scalar-tensor theories and obtain theory-independent inequalities.
After discussing the mass definition in scalar-tensor theories,
these inequalities are 
related to a theory-dependent maximum mass-to-size ratio. 
We show
that its value can exceed not only Buchdahl's limit, 8/9, but also unity, 
which we call {\it the black hole limit}, 
in contrast to general relativity.
Next, we numerically examine the validity of the assumptions made in
deriving the inequalities and the applicability of our analytic results.
We find that the assumptions are mostly satisfied and that the mass-to-size 
ratio exceeds both Buchdahl's limit and the black hole limit.
However, we also find that this ratio never exceeds Buchdahl's limit
when we impose the further condition, $\rho-3p\ge0$, 
on the density, $\rho$, and pressure, $p$, of the matter. 
\end{abstract}

\

\section{Introduction}

\

Einstein's general relativity postulates that gravitational interactions
are mediated by a tensor field, $g_{\mu\nu}$.
It is also well-known that 
electro-magnetic interactions are mediated by a 
vector field, $A_{\mu}$. 
One may therefore suspect 
that some unknown interactions may be mediated by scalar fields.
Such theories have been suggested since before the appearance of 
general relativity.
Moreover,
it has been repeatedly pointed out over the years 
that unified theories that contain gravity as well as other
interactions  
naturally give rise to scalar fields
coupled to matter with gravitational strength.  
This motivation has led many theoretical physicists to study scalar-tensor 
theories of gravity (scalar-tensor theories)
\cite{4},\cite{15},\cite{14},\cite{5}.
The scalar-tensor theories are natural alternatives to general relativity,
and gravity is mediated not only by a tensor field 
but also by a scalar field in these theories. 
Recently, such theories have been of interest
as effective theories of string theory at low energy scales \cite{16}.

Many predictions of the scalar-tensor theories in 
 strong gravitational fields are
summarized in Ref.\cite{5},\cite{2},\cite{3}. 
It has been found that a wide class of scalar-tensor theories can pass 
all experimental tests in weak gravitational fields.
However, it has also been found that scalar-tensor theories
exhibit different aspects of gravity in
strong gravitational fields in contrast to general relativity.
It has been shown numerically 
that nonperturbative effects in the scalar-tensor 
theories increase the maximum mass of an isolated system
such as a neutron star\cite{2},\cite{3}.  
 
In general relativity, 
the mass-to-size ratio of a star
has physical significance, especially for an isolated system.
Buchdahl has obtained a maximum value of
the mass-to-size ratio of a static and spherically symmetric star 
under the following physically reasonable 
assumptions \cite{11},\cite{7},\cite{6}.
\begin{itemize}
\item No black hole exists.
\item The constitution of the star is a perfect fluid.
\item The density at any point in the star is a positive and 
 monotonously decreasing 
function of the radius.
\item An interior solution of the star smoothly matches an exterior solution,
i.e., Schwarzschild's solution.
\end{itemize}
Buchdahl has obtained an upper limit of the mass-to-size ratio
as $2M/R\leq8/9$. We shall refer to this as the  
{\it Buchdahl inequality}.

Motivated by Buchdahl's theorem,
we shall derive a modified Buchdahl inequality 
to obtain the maximum mass-to-size ratio
in scalar-tensor theories. 
We then numerically examine
the validity of the assumptions made in deriving the inequality.
The applicability of our analytic results is also examined.
This paper is organized as follows.
In section 2, we summarize the basic equations in the scalar-tensor theories.
In section 3, we derive a modified
Buchdahl inequality in the scalar-tensor theories,
and the numerical results are compared with the analytic 
results in section 4.
A brief summary is given in section 5.   

\

\section{Basic equations}

\

We shall consider the simplest 
scalar-tensor theory \cite{4},\cite{5},\cite{8}.
In this theory, gravitational interactions are mediated by 
a tensor field, $ g_{\mu\nu} $, and a scalar field, $ \phi $.
The action of the theory is 
\begin{equation}
S=\frac{1}{16\pi}\int\sqrt{-g}\left[\phi R-\frac{\omega(\phi)}
{\phi}g^{\mu\nu}\phi_{,\mu}\phi_{,\nu}\right]d^4x + 
S_{matter}[\Psi_{m},g_{\mu\nu}],
\label{e5}
\end{equation}
where $\omega(\phi)$ is a dimensionless arbitrary function of $\phi$,
$\Psi_m$ represents matter fields, and $S_{matter}$
is the action of the matter fields.
The scalar field, $\phi$, plays the role of an effective
gravitational constant as $G\sim1/\phi$.
Varying the action by the tensor field, $g_{\mu\nu}$, 
and the scalar field, $\phi$, yields, respectively,
the following field equations:
\begin{eqnarray}
G_{\mu\nu}& = & \frac{8\pi}{\phi}T_{\mu\nu}+\frac{\omega(\phi)}{\phi^2}
\left(\phi_{,\mu}\phi_{,\nu}-\frac{1}{2}g_{\mu\nu}g^{\alpha\beta}\phi_{,\alpha}
\phi_{,\beta}\right)+\frac{1}{\phi}(\nabla_{\mu}\nabla_{\nu}\phi-g_{\mu\nu}
\Box\phi), \label{e6} \\
\Box\phi & = & \frac{1}{3+2\omega(\phi)}\left(8\pi T - \frac{d\omega}{d\phi}
g^{\alpha\beta}\phi_{,\alpha}\phi_{,\beta}\right).
\label{e7}
\end{eqnarray}

Now we perform the conformal transformation, 
\begin{equation}
g_{\mu\nu}=A^2(\varphi)g_{\ast\mu\nu},
\label{e8}
\end{equation}
such that
\begin{equation}
G_{\ast}A^2(\varphi) = \frac{1}{\phi}, 
\label{e9}
\end{equation}
where $G_{\ast}$ is a bare gravitational constant, and 
we call $A(\varphi)$ {\it{a coupling function}}.
Hereafter, the symbol, $\ast$, denotes quantities or derivatives associated
with $g_{\ast\mu\nu}$.
Then the action can be rewritten as
\begin{equation}
S=\frac{1}{16\pi G_{\ast}}\int\sqrt{-g_{\ast}}(R_{\ast}
-2g_{\ast}^{\mu\nu}\varphi_{,\mu}\varphi_{,\nu}) d^4x
+ S_{matter}[\Psi_{m},A^2(\varphi)g_{\ast\mu\nu}],
\label{e10}
\end{equation}
where the scalar field, $\varphi$, is defined by
\begin{equation}
\alpha^2(\varphi) \equiv \left(\frac{d\ln A(\varphi)}{d\varphi}\right)^2
= \frac{1}{3+2\omega(\phi)}. 
\label{e11}
\end{equation}
Varying the action by $g_{\ast\mu\nu}$ and $\varphi$ 
yields, respectively, 
\begin{eqnarray}
G_{\ast\mu\nu}& = & 8\pi G_{\ast} T_{\ast\mu\nu}
+ 2\left(\varphi_{,\mu}\varphi_{,\nu}
-\frac{1}{2}g_{\ast\mu\nu}g_{\ast}^{\alpha\beta}\varphi_{,\alpha}
\varphi_{,\beta}\right),
\label{e12} \\
\Box_{\ast}\varphi & = &-4\pi G_{\ast}\alpha(\varphi) T_{\ast}, 
\label{fs2}
\end{eqnarray}
where $T_{\ast}^{\mu\nu}$ represents the energy-momentum tensor 
with respect to $g_{\ast\mu\nu}$ defined by
\begin{equation}
T_{\ast}^{\mu\nu} \equiv \frac{2}{\sqrt{-g_{\ast}}}
\frac{\delta S_{matter}[\Psi_{m},A^2(\varphi)g_{\ast\mu\nu}]}
{\delta g_{\ast\mu\nu}}=A^6(\varphi)T^{\mu\nu}.
\label{e14}
\end{equation}
The conservation law for $T_{\ast}^{\mu\nu}$ is given by
\begin{equation}
\nabla_{\ast\nu}T_{\ast\mu}^{\nu}=\alpha(\varphi)T_{\ast}
\nabla_{\ast\mu}\varphi.
\label{e15}
\end{equation}

The field equations $(\ref{fs2})$ and $(\ref{e15})$ tell us that  
the {\it coupling strength}, $\alpha(\varphi)$, 
plays a role in mediating interactions between 
the scalar field, $\varphi$, and the matter.
General relativity is characterized by having a vanishing coupling strength:
$\alpha(\varphi)=0$, i.e., $A(\varphi)=1$.
The Jordan-Fierz-Brans-Dicke theory is characterized by having a 
$\varphi$-independent couping strength: 
$\alpha(\varphi)=\alpha_0=\mbox{const}.$, 
i.e., $A(\varphi)=e^{\alpha_0 \varphi}$ \cite{4},\cite{2}.
Observational constraints on the coupling strength are
summarized in Appendix A.

\

\section{Modified Buchdahl's theorem in scalar-tensor theories} 

\

In this section, we consider a static and spherically 
symmetric space-time with a perfect fluid.
First, we derive a modified Buchdahl inequality.
Then the inequality is reformulated to obtain
the maximum value of the mass-to-size ratio
in the scalar-tensor theories. 
Hereafter, we refer to $(g_{\mu\nu},\phi)$ 
and $(g_{\ast\mu\nu},\varphi)$, respectively, as 
the physical frame and the Einstein frame.

\

\subsection{A modified Buchdahl inequality in scalar-tensor theories}

\

In the Einstein frame, a line element of 
the static and spherically symmetric space-time is written as \cite{6}
\begin{equation}
ds^2_{\ast}=-f_{\ast}(r)dt^2+h_{\ast}(r)dr^2+r^2d\Omega^2.
\label{e57}
\end{equation}
The stress-energy tensor for the perfect fluid in the  
Einstein frame is given by
\begin{subequations}
\begin{equation}
T_{\ast}^{\mu\nu}=(\rho_{\ast}+p_{\ast})
u_{\ast}^{\mu}u_{\ast}^{\nu}+p_{\ast}g_{\ast}^{\mu\nu},
~~u_{\ast\alpha}=-\sqrt{f_{\ast}(r)}~(dt)_\alpha,
\label{e58}
\end{equation}
where $u_{\ast}^{\alpha}$ is the four velocity of the matter.
The stress-energy tensor for the perfect fluid in the   
physical frame is given by
\begin{equation}
T^{\mu\nu}=(\rho +p)
u^{\mu}u^{\nu}+pg^{\mu\nu},
~~u_{\alpha}=-A\sqrt{f_{\ast}(r)}~(dt)_\alpha,
\end{equation}
\end{subequations}
where $u^{\alpha}$ is the four velocity of the matter.
Fluid variables in the physical and Einstein frames
are related according to 
\begin{eqnarray}
u^{\alpha} & = & u_{\ast}^{\alpha}
A^{-1}(\varphi), \label{e60} \\
\rho & = & \rho_{\ast}A^{-4}(\varphi), \label{je74} \\
p & = & p_{\ast}A^{-4}(\varphi). \label{je75}  
\end{eqnarray}
Now the field equations $(\ref{e12})$ and
$(\ref{fs2})$ are reduced to the following equations:
\begin{eqnarray}
\left(r(1-h_{\ast}^{-1})\right)' & = &
8\pi G_{\ast}\rho_{\ast}r^2+\frac{r^2}{h_{\ast}}\varphi^{'2}, 
\label{f1} \\
-r^{-2}h_\ast(1-h_\ast^{-1})+r^{-1}f_\ast^{-1}{f_\ast}' & = & 
8\pi G_{\ast}p_{\ast}h_\ast+\varphi^{'2}, 
\label{f2} \\
\left(\frac{f_{\ast}^{'}}{2f_{\ast}}\right)^{'}
+\left(\frac{f'_{\ast}}{2f_{\ast}}\right)^2
+\frac{1}{r}\frac{f'_{\ast}}{2f_{\ast}}
-\frac{h'_{\ast}}{2h_{\ast}}\frac{f'_{\ast}}{2f_{\ast}}
-\frac{1}{r}\frac{h'_{\ast}}{2h_{\ast}} & = &
8\pi G_{\ast}p_{\ast}h_{\ast}-\varphi^{'2}, 
\label{f3} \\
r^{-2}(f_{\ast}h_{\ast})^{-\frac{1}{2}}
\left(\left(\frac{f_{\ast}}
{h_{\ast}}\right)^{\frac{1}{2}}r^2\varphi'\right)' & = &
4\pi G_{\ast}\alpha(\varphi)(\rho_{\ast}-3p_{\ast}), \label{f4} 
\end{eqnarray}
where a prime denotes differentiation with respect to $r$.
As is often done in the cases of general relativity,
we define a {\it mass function}, $m_\ast(r)$, 
in the Einstein frame as follows: 
\begin{equation}
h_{\ast}(r) \equiv \left[1-\frac{2m_{\ast}(r)}{r}\right]^{-1}.
\label{e61}
\end{equation}
Then $(\ref{f1})$ is rewritten as
\begin{equation}
m_{\ast}^{'}(r) = 4\pi G_{\ast}\re r^2,  \label{f1a} 
\end{equation}
where 
\begin{equation}
\re \equiv \rho_{\ast}+\frac{\varphi^{'2}}
{8\pi G_{\ast}h_{\ast}}.   \label{eff}
\end{equation}
That is, $\re$ plays the role of an 
effective density in the Einstein frame.

In order to derive a modified Buchdahl inequality,
we assume 
\begin{eqnarray}
h_{\ast}(r) & \ge & 0,                    \label{as1} \\
f_{\ast}(r) & \ge & 0,                    \label{as2} \\
\re(r) & \ge & 0,           \label{as3} \\
\red(r) & \le & 0.       \label{as4}
\end{eqnarray}
These assumptions imply the following: 
\begin{itemize}
\item No black hole exists in the Einstein frame. 
\item The effective density, $\re$, is a positive and monotonously 
decreasing function of the radius.
\end{itemize}
Moreover, we assume that
an interior solution of the above field equations 
smoothly matches the corresponding exterior one.
Note that these assumptions are concerned with 
the unphysical variables and that
their validity should be examined. 
This will be done later.
 
Using the assumption $(\ref{as4})$, it is easy to verify
the following inequality:
\begin{equation}
\left(\frac{m_\ast}{r^3}\right)'\leq0.
\label{mine}
\end{equation}
Moreover, with $(\ref{f2})$, $(\ref{f3})$ and $(\ref{f1a})$,
we obtain 
\begin{equation}
-\left(\frac{\left(\sqrt{f_{\ast}}\right)'}{r\sqrt{h_{\ast}}}\right)^{'}
=\sqrt{f_{\ast}h_{\ast}}\left(-\left(\frac{m_{\ast}}{r^3}\right)'
+2\frac{\varphi^{'2}}{rh_{\ast}}\right) \ge 0. 
\label{b1}
\end{equation}
Accordingly, we have
\begin{equation}
\frac{\left(\sqrt{f_{\ast}(r_1)}\right)'}
{r_1\sqrt{h_{\ast}(r_1)}}
\geq
\frac{\left(\sqrt{f_{\ast}(r_2)}\right)'}
{r_2\sqrt{h_{\ast}(r_2)}},~~~r_1\leq r_2.
\label{add}
\end{equation}

Now let the inequality $(\ref{add})$ be reformulated 
in terms of variables of the exterior solution.
The exterior solution, whose derivation is given in Appendix B, 
is 
\begin{equation}
{ds_\ast}^2=-e^{\gamma(\chi)}dt^2+e^{-\gamma(\chi)}d\chi^2+
e^{\lambda(\chi)-\gamma(\chi)}d\Omega^2,
\label{add2}
\end{equation}
where
\begin{eqnarray}
e^{\lambda(\chi)} & = & \chi^2-a\chi, \label{je30} \\
e^{\gamma(\chi)} & = & \left(1-\frac{a}{\chi}\right)^{\frac{b}{a}}, 
\label{je31} \\
\varphi(\chi) & = & \varphi_0 + \frac{c}{a}
\ln\left(1-\frac{a}{\chi}\right). \label{je32}
\end{eqnarray}
Here $a$, $b$, $c$ and $\varphi_0$ are constants of integration,
and $\varphi_0$ is the asymptotic value of $\varphi$ at infinity.
Moreover, the constants, $a$, $b$ and $c$, must satisfy
the following relation (Appendix B): 
\begin{equation}
a^2-b^2 = 4c^2.
\label{jconst} 
\end{equation}

One may expect that $\chi=a$ is an event horizon. 
However,
this is not the case in generic scalar-tensor theories,
where the null surface, $\chi=a$, is a curvature singularity in the Einstein frame. 
The singular nature of the unphysical space-time at $\chi=a$
can also be seen
when transformation to the Schwarzschild coordinate
is made. 
The Schwarzschild coordinate, $r$, and the Just coordinate, $\chi$,
are related by the following relation:
\begin{equation}
r=\chi\left(1-\frac{a}{\chi}\right)^{\frac{a-b}{2a}}.
\label{jadd2}
\end{equation}
One finds that, when $a\neq b$, $\chi=a$ in the Just coordinate
corresponds to $r=0$ in the Schwarzschild coordinate.

By matching the interior solution to the exterior solution, 
we obtain the following relations:
\begin{eqnarray}
r_s & = & \chi_s 
\left( 1-\frac{a}{\chi_s}\right)^{\frac{a-b}{2a}}, \label{e62} \\
f_{\ast}(r_s) & = & \left(1-\frac{a}{\chi_s}\right)^{\frac{b}{a}}, \label{e63} \\
h_{\ast}(r_s) & = & \left(1-\frac{a}{\chi_s}\right)
\left(1-\frac{a+b}{2\chi_s}\right)^{-2}, 
\label{b4} 
\end{eqnarray}
where the subscript, $s$, refers to $\chi$ evaluated 
at the surface, $r=r_s$.
Note that 
\begin{equation}
\chi_s > a\ge b.\label{add3}
\end{equation}
Since ${h_\ast(r)}^{-1}=1-2m_\ast(r)/r$,
$(\ref{b4})$ becomes
\begin{equation}
2m_{\ast}(r_s) =
\left(b-\frac{(a+b)^2}{4\chi_s}\right)
\left(1-\frac{a}{\chi_s}\right)^{-\frac{(a+b)}{2a}}.
\label{b7} 
\end{equation}
Accordingly, by virtue of the positivity of $m_\ast(r)$,
we obtain the additional inequality
\begin{equation}
b-\frac{(a+b)^2}{4\chi_s} > 0.
\label{b6}
\end{equation}
With $(\ref{add})$ and $(3.25)\sim(3.28)$,  
we obtain the following relation for $r\le r_s$: 
\begin{equation}
\frac{\left(\sqrt{f_{\ast}}\right)'}{r\sqrt{h_{\ast}}}
\ge \frac{\left(\sqrt{f_{\ast}(r_s)}\right)'}{r_s\sqrt{h_{\ast}(r_s)}}
= \frac{b}{2{r_s}^3}.
\label{b2}
\end{equation}
Integrating $(\ref{b2})$ from the center, $r=0$, to the surface, $r=r_s$, 
we obtain 
\begin{eqnarray}
0 & \le & \sqrt{f_{\ast}(0)} \nonumber \\
& \le & 
\sqrt{f_{\ast}(r_s)}-\frac{b}{2{r_s}^3}
\int_{0}^{r_s}r\sqrt{h_{\ast}(r)}dr \nonumber \\
& \le & 
\sqrt{f_{\ast}(r_s)}-\frac{b}{2{r_s}^3}\int_{0}^{r_s}
r\left(1-\frac{2m_{\ast}(r_s)}{{r_s}^3}r^2\right)^{-\frac{1}{2}}
dr \nonumber \\
& = & 
\left(1-\frac{a}{\chi_s}\right)^{\frac{b}{2a}}+
\frac{b}{4m_{\ast}(r_s)}\left(\sqrt{1-\frac{2m_{\ast}(r_s)}{r_s}}-1\right),
\label{b3}
\end{eqnarray}
where the inequality $(\ref{mine})$ has been used.

The inequalities obtained to this point can be simplified 
in terms of new parameters defined by
\begin{equation}
a_s \equiv \frac{a}{\chi_s}, \ \ \ \ 
b_s \equiv \frac{b}{\chi_s}, \ \ \ \ 
c_s \equiv \frac{c}{\chi_s}. 
\label{ns} 
\end{equation}
Substituting $(\ref{b7})$ into $(\ref{b3})$, 
we obtain the following inequality:
\begin{eqnarray}
0 & \le  & \sqrt{f_{\ast}(0)} \nonumber \\ 
  & \le  & 
\frac{1}{2}(1-a_s)^{\frac{b_s}{2a_s}}
\left(b_s-\frac{(a_s+b_s)^2}{4}\right)^{-1}\times \nonumber \\
& & \times
\left[2b_s-\frac{(a_s+b_s)^2}{2}-b_s\sqrt{1-a_s}
\left(1-\sqrt{1-\frac{4b_s-(a_s+b_s)^2}{4(1-a_s)}}\right)\right].
\label{e65}
\end{eqnarray}
The above inequality can be further simplified,  
and we finally obtain a modified Buchdahl inequality
in the scalar-tensor theories as 
\begin{equation}
F(a_s,b_s)\equiv 3b_s-\frac{1}{2}(a_s+b_s)(a_s+2b_s)-b_s\sqrt{1-a_s}
\ge 0,
\label{ke1} 
\end{equation}
supplemented with $(\ref{add3})$ and $(\ref{b6})$.

\begin{figure}
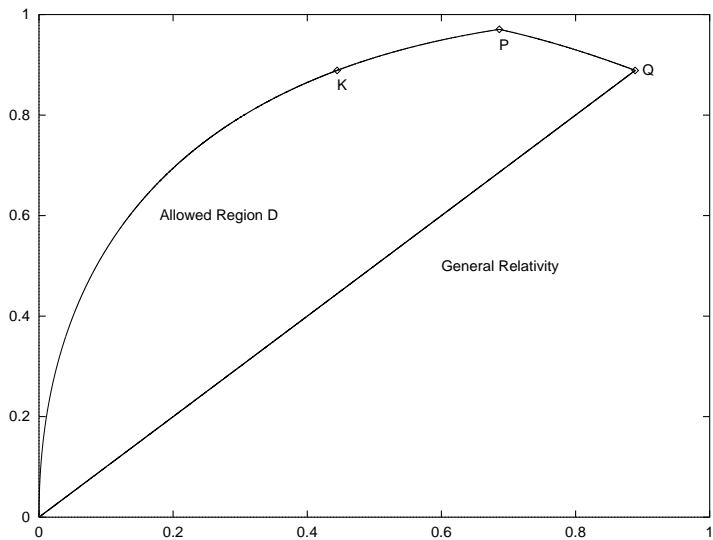

\begin{center}
\hspace*{0.5cm}
\end{center}
\caption{The allowed region, $D$, is depicted,
where horizontal and vertical axes denote, respectively,
$b_s$ and $a_s$.
  The characteristic points,
  $P=(4(3-2\sqrt{2}),4(3\sqrt{2}-4))$, $Q=(8/9,8/9)$ and
  $K=(4/9,8/9)$, are shown.
  In general relativity, $a_s=b_s$.
  Buchdahl's limit is denoted by $Q$.
  On $K$, $|c_s|$ takes the maximum value,
  $|c_s|_{\mbox{\tiny MAX}}=2\sqrt{3}/9$. }
\label{reg}
\end{figure}

The modified Buchdahl inequality can be solved to yield  
\begin{equation}
\left.
\begin{array}{lcl}
b_s \le a_s \le 2\sqrt{b_s}-b_s & \mbox{for} & 0 \le b_s \le 4(3-2\sqrt{2}),
\\
b_s \le a_s \le 2\sqrt{2b_s}-2b_s & \mbox{for} & 4(3-2\sqrt{2}) \le b_s \le
\frac{8}{9}, \\
\mbox{Forbidden}  & \mbox{for} & b_s > \frac{8}{9}. 
\end{array}
\right\}
\label{e74}
\end{equation}
Fig. \ref{reg} displays the allowed region, $D$, of $(a_s,b_s)$. 

In addition, we can obtain, with $(\ref{jconst})$,
the following upper limit on $|c_s|$:
\begin{equation}
|c_s| \le \frac{2\sqrt{3}}{9}.
\label{e75}
\end{equation}

The inequality $(\ref{e74})$ is significant.
The third inequality of $(\ref{e74})$ gives us a necessary condition
for a spherical star to exist 
and is reduced to Buchdahl's theorem in general relativity 
when $c=0$, i.e., $a=b$, and, accordingly, $\chi=r$.
In this case, we have ($R=r_s$)
\begin{equation}
c_s=0 \Longleftrightarrow a_s = b_s = \frac{2M}{R} \le \frac{8}{9},
\label{buchg}
\end{equation}
where $M$ is the ADM mass defined at spatial infinity,
and $R$ is related to the surface area, $S$, as $S=4\pi R^2$.

The new and important inequality $(\ref{e75})$ 
is characteristic of the scalar-tensor
theories and does not have a general relativistic counterpart. 
It has been found the appearance of 
nonperturbative behavior of the scalar
field in the previous numerical studies \cite{2},\cite{3}. 
Our result implies that, even in a strong gravitational field,
the characteristic amplitude of the scalar field, $|c_s|$, is bounded. 

It is important to note that we have not used any assumption
regarding the coupling function, $A(\varphi)$, in deriving
the inequalities. In particular, the inequalities
$(\ref{e74})$ and $(\ref{e75})$ give theory-independent
constraints on the parameters, $b_s$ and $c_s$.

\

\subsection{The mass-to-size ratio}

\

Now we reformulate the inequalities derived in the previous section,
which are in terms of variables in the Einstein frame,
in order to obtain the mass-to-size ratio in the physical frame.
To do this, the coupling function, $A(\varphi)$, should be
specified.
In this paper, we assume as an example of this coupling function 
the simple form 
\begin{equation}
A(\varphi) = e^{\frac{1}{2}\beta\varphi^2},
\label{cup}
\end{equation}
where $\beta$ is a constant\cite{2},\cite{3}.
Then the coupling strength, $\alpha(\varphi)$, becomes
\begin{equation}
\alpha(\varphi)=\beta\varphi.
\label{jalpha}
\end{equation}

A natural definition of the radius of a spherical star
is obtained by using its (physical) surface area as follows.
In the physical frame, the surface area, $S$,
is given by
\begin{eqnarray}
S & = & 4\pi A^2(\varphi_s)e^{\lambda(\chi_s)-\gamma(\chi_s)} \nonumber \\
  & = & 4\pi\chi_s^2(1-a_s)^{1-\frac{b_s}{a_s}}
\exp\left[\beta\left(\frac{c_s}{a_s}\ln(1-a_s)\right)^2\right], 
\label{area}
\end{eqnarray}
where we take 
the asymptotic value of the scalar field as $\varphi_0=0$, and,
accordingly, we have $A(\varphi_0)=1$ and 
$\alpha(\varphi_0)$=0.
This surface area defines the physical radius, $R$, of the star 
in a similar manner as in general relativity:
\begin{equation}
R \equiv \sqrt{\frac{S}{4\pi}}. \label{e70}
\end{equation}
When $\varphi_0=0$, 
the effective gravitational constant, $G$, defined in Appendix A,
is equal to $G_{\ast}$.
If $\varphi_0 \neq 0$, contributions of the scalar field  
appear in the above expression of $R$ in terms of $c\,\beta\,\varphi_0$.


The definition of the mass in the Brans-Dicke theory
is found in Ref.\cite{20}, and the mass in the scalar-tensor
theories is defined in the same manner.
In defining the mass, the metric should be expressed in 
the isotropic coordinate, $\bar{r}$, which
is related to $\chi$ as
\begin{equation}
\chi=\bar{r}\left(1+\frac{a}{4\bar{r}}\right)^2.
\label{iso1}
\end{equation}
In our model,
the exterior solution is then rewritten in terms of $\bar{r}$ as
\begin{equation}
\eqalign{
&ds^2  =  A^2(\varphi)\left[
-\left(\frac{1-\frac{a}{4\bar{r}}}{1+\frac{a}{4\bar{r}}}\right)^{\frac{2b}{a}}
dt^2 +
\left(1+\frac{a}{4\bar{r}}\right)^{\frac{2(a+b)}{a}}
\left(1-\frac{a}{4\bar{r}}\right)^{\frac{2(a-b)}{a}}
\left(d\bar{r}^2+\bar{r}^2 d\Omega^2\right)\right] \cr
&G_\ast\phi  =  A^{-2}(\varphi)
=\exp\left[-\frac{4\beta c^2}{a^2}\left(
\ln\frac{1-\frac{a}{4\bar{r}}}{1+\frac{a}{4\bar{r}}}
\right)^2\right].\cr}\label{is3} 
\end{equation}
By introducing the asymptotic Cartesian coordinates
such that $\bar{r}=\sqrt{(x^1)^2+(x^2)^2+(x^3)^2}$,
the asymptoric form of the solution is easily found to be
\begin{equation}
\eqalign{
&G_\ast\phi \sim 1 + \frac{0}{\bar{r}} 
\equiv 1 + \frac{2M_S}{\bar{r}},~~~~~
g_{00} \sim -1 + \frac{b}{\bar{r}}\equiv -1 + 
\frac{2(M_T+M_S)}{\bar{r}},\cr
&g_{ij} \sim \left(1 + \frac{b}{\bar{r}}\right)\delta_{ij}
\equiv \left(1 + \frac{2(M_T-M_S)}{\bar{r}}\right)\delta_{ij},\cr
}
\label{is4}
\end{equation}
where the quantities, $M_S$ and $M_T$, 
are called, respectively,
the scalar mass and the tensor mass. 
At Newtonian order, their sum, $M\equiv M_T+M_S$,
plays the role of the mass and is called the active gravitational mass. 
In our model, $M_S=0$, and, accordingly, $M_T=M=b/2$.
Hereafter, we call $M$ the mass for simplicity.

\begin{figure}
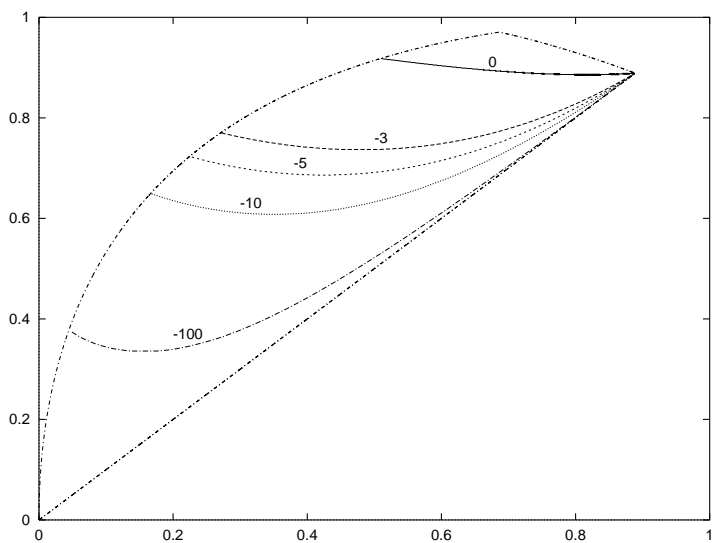

\begin{center}
\hspace*{0.5cm}
\caption{On each lines, $H(a_s,b_s;\, \beta)=8/9$ for
  various values of $\beta$: $0,-3,-5,-10$ and $-100$.
  Horizontal and vertical axes denote, respectively, $b_s$ and $a_s$.}
\label{fig1}
\end{center}
\end{figure}

\begin{figure}
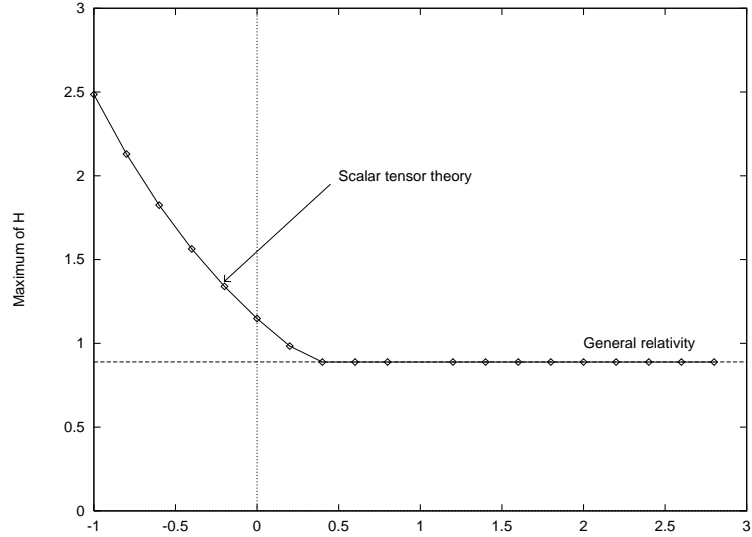

\begin{center}
\hspace*{0.5cm}
\caption{The Maximum mass-to-size ratio, $H_{\mbox{\tiny MAX}}$,
  is shown as a function of $\beta$.
  Horizontal and vertical axes denote, respectively, $\beta$,
  and $H_{\mbox{\tiny MAX}}$.
  A horizontal line, $H_{\mbox{\tiny MAX}}=8/9$,
  denotes Buchdahl's limit in general relativity.
  When $\beta \lesssim 0.4$, $H_{\mbox{\tiny MAX}}$ exceeds
  Buchdahl's limit.
  When $\beta\lesssim0.2$, $H_{\mbox{\tiny MAX}}$
  exceeds unity, a black hole limit.}
\label{ff}
\end{center}
\end{figure}

Now we are ready to calculate the mass-to-size ratio
in the scalar-tensor theory
as a function of $a_s$, $b_s$, $\chi_s$ and 
a specific parameter of our model, $\beta$. We obtain
\begin{equation}
H(a_s,b_s;\, \beta) \equiv
\frac{b}{R}
= b_s(1-a_s)^{\frac{b_s-a_s}{2a_s}}
\exp\left[-\frac{1}{2}\beta
\left(\frac{c_s}{a_s}\ln(1-a_s)\right)^2\right]. \label{e73}
\end{equation}
In Fig.\ref{fig1}, 
in the allowed region of $(a_s,b_s)$, we display lines  
on which $H(a_s,b_s;\, \beta)$ is equal to $8/9$ 
for various values of $\beta$.
For a fixed value of $\beta$, the region above the line
corresponds to the case that the mass-to-size ratio, 
$2M/R$,
exceeds Buchdahl's limit, 8/9. 
Moreover, in some cases, 
it may be greater than unity. 
We refer to this case as the  
{\it black hole limit}.
The maximum values of $H(a_s,b_s;\, \beta)$ for various values of $\beta$ 
are shown in Fig.\ref{ff}.
Indeed, the maximum mass-to-size ratio can sometimes become larger
than the black hole limit.
However, 
the physical exterior solution generically 
does not have an event horizon in scalar-tensor theories,
in contrast to general relativity, and, therefore, the condition,
$2M/R>1$, does not imply the existence of a black hole. 

Now suppose that a space rocket approaches a star for which $2M/R>1$ and
goes into its {\it Schwarzschild radius} defined by $2M$.
A spaceman in the rocket would be resigned to his fate to die,
but we know that he still has a chance 
to return alive from a {\it false black hole}.


\section{Numerical results}

\

Equations $(\ref{f1})\sim(\ref{f4})$ are numerically solved to
obtain an interior solution.
This solution is then matched to the exterior one,
and numerical values of the parameters $a_s$, $b_s$ and $c_s$
are calculated.
Some details of the numerical methods are
summarized in Appendix C.
Since we take $\varphi_0=0$,
$G_{\ast}$ is equal to $G$ (Appendix A).
Hereafter, we use units in which $G_{\ast}=G=1$.

As for the matter,
we assume the following polytropic equation of state \cite{13}:
\begin{eqnarray}
\rho & = & m_b n + \frac{Kn_0m_b}{\Gamma-1}
\left(\frac{n}{n_0}\right)^{\Gamma}, \label{eos1} \\
p & = & Km_bn_0\left(\frac{n}{n_0}\right)^{\Gamma}, \label{eos2} \\
m_b & = & 1.66\times 10^{-24}\mbox{g} \label{eos3}, \\
n_0 & = & 0.1\mbox{fm}^{-3}. \label{eos4}
\end{eqnarray}
We take the parameters values, $\Gamma = 2.34 $ and $K=0.0195$ \cite{2},
which fit a realistic equation of state of high density
nuclear matter, and probably also that of a neutron star quite well.
Our numerical solutions are therefore parametrized by
$\beta$ and $n_c\equiv n(0)$.
It has been shown numerically that significant effects of
$\varphi$ appear when $\beta\leq-4.35$ \cite{2},\cite{3},
and we are mostly interested in cases of negative values of
$\beta$. In cases of positive values of $\beta$,
we cannot numerically find any significantly different behavior
of the solutions compared with those in general relativity,
and any further discussion in these cases is no longer done.

\begin{figure}
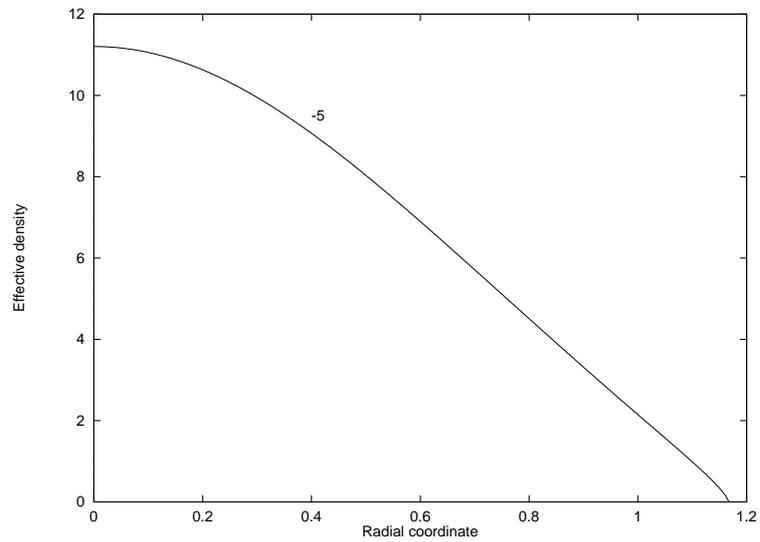

\begin{center}
\hspace*{0.5cm}
\caption{The effective density, $\re(r)$, is shown
in the case that $\beta=-5$ and $n_c/n_0=10$.
Horizontal and vertical axes denote, respectively,
the radial coordinate, $r$, in the unit of $10\mbox{km}$,
and the effective density, $\re(r)/(m_bn_0)$.
It is seen that the assumption, $\red(r) \le 0$, is satisfied.}
\label{f-5.1}
\end{center}
\end{figure}

First, we examine whether our assumption,
$\re'\leq0$, is satisfied.
In Fig.\ref{f-5.1}, we give an example of numerical behavior of
the effective density for $\beta=-5$ and
$n_c/n_0=10$.
Including this case, we find that
the assumption, $\red(r) \le 0$, is mostly satisfied, 
as summarized in the 3rd column of Table 1.
Differentiating $(\ref{eff})$, we obtain
\begin{equation}
\red=A^4(\varphi)\rho' +
4A^4(\varphi)\alpha(\varphi)\varphi'\rho+\left(\frac{{\varphi'}^2}{8\pi
    G_{\ast}h_{\ast}}\right)'.
\label{abc}
\end{equation}
In Fig.\ref{f-5.2}, we show the 1st, 2nd and 3rd terms in
$(\ref{abc})$
with the same parameters as Fig.\ref{f-5.1}.
The first term is indeed dominant and always negative.
In Fig.\ref{f-5.4},
we compare the corresponding physical quantities,
$\rho(r)$ and $\phi(r)$, and the
unphysical scalar field, $\varphi(r)$.
It is found that the {\it local gravitational constant},
$G(\phi)\equiv1/\phi$, increases as
$\rho$ decreases toward the surface. However, this behavior
is strongly dependent on the coupling function, and
the sign of $\beta$ is crucial in the present case.

Next, we examine an extreme example
in which the assumption, $\re'\leq0$, is violated.
Fig.\ref{fb} shows the effective density, $\re$,
for $\beta=-30$ and $n_c/n_0=10$.
It is seen that $\re$ remains constant in the central part
and then increases between the two rectangles in Fig.\ref{fb}.
In Fig.\ref{fdif}, we show the three terms in $(\ref{abc})$ and find
that the positive second term becomes partially dominant.
Again, this behavior
is strongly dependent on the coupling function.
In the present case, we have $\alpha(\varphi)=\beta\varphi$,
where $\beta<0$. Therefore, when $\varphi'<0$ and $|\beta|$
is large, 
such that the second term in $(\ref{abc})$ is dominant,
$\re'$ becomes positive, and the assumption is violated.
In Fig.\ref{f-30.cl}, we compare $\rho(r)$, $\phi(r)$ and
$\varphi(r)$. It is found that, despite small values of
$\varphi$, $\phi$ can be large due to a large value of $|\beta|$.
However, it should be noted that this assumption
is concerned with the unphysical quantity, $\re$,
and that its violation does not necessarily mean
that this extreme case is unreal.
In Fig.\ref{fc}, we show the energy density, $\rho(r)$, and the
pressure, $p(r)$, in the physical frame in the extreme case:
$\beta=-30$, $n_c/n_0=10$.
The behavior of these quantities seems ordinary, that is,
they are monotonously decreasing functions of $r$.
Accordingly, one may think that
this can be a physically acceptable equilibrium solution
despite the violation of the assumption, $\red\le0$.
However, we are forbidden to take $\beta$ smaller than $-5$
because of experimental
constraints (Appendix A, \cite{3},\cite{10}).

For each value of $\beta$, the
mass-to-size ratio, $2M/R=H(a_s,b_s;\beta)$,
can be numerically calculated as a function of $n_c$.
By changing $n_c$, we search for a maximum value of
$H(a_s,b_s;\,\beta)$ for each $\beta$.
In the 5th column of Table\ref{t1},
we summarize our results for the maximum mass-to-size
ratio, where the parameters are chosen such that
the assumption, $\red \le 0$, is satisfied.
For $\beta<-12.07$, we find numerically that
the assumption is always violated.
The first interesting example is found in the case, $\beta=-12.07$,
in which the maximal mass-to-size ratio is obtained
as $H_{\mbox{\tiny MAX}}=1.018$ when $n_c/n_0=11.2$.
This is a case in which 
$H_{\mbox{\tiny MAX}}$ exceeds the black hole limit, $H=1$.
Another interesting example is found in the case, $\beta=-11$,
in which $H_{\mbox{\tiny MAX}}=0.919$ when $n_c/n_0=11.3$.
This is a case in which 
$H_{\mbox{\tiny MAX}}$ exceeds Buchdahl's limit, $H=8/9\approx0.889$.
These examples have academic importance in the sense
that our analytic results in the previous section
are partially realized also in the numerical solutions.
 
To this point, the stability of our numerical solutions has not been taken 
into account. We have found that $\rho-3p$ may be a good estimator
of the stability, as described below.
The baryonic mass of a star is defined as \cite{2}
\begin{equation}
\bar{m}=
m_b\int^{r_s}_{0}4\pi nA^3(\varphi)r^2
\left(1-\frac{2m_{\ast}}{r}\right)^{-\frac{1}{2}}dr.
\end{equation} 
We have numerically examined how $\bar{m}$ depends on
$n_c$ and find a significant correlation between
the signature of $d\bar{m}/dn_c$ and that of $\rho-3p$.
That is, the cases, $\rho-3p < 0$ and $\rho-3p > 0$, approximately 
correspond to the cases, $d\bar{m}/dn_c < 0$ and $d\bar{m}/dn_c > 0$,
respectively.
Accordingly, we shall interpret the signature of $\rho-3p$
as a measure of the onset of the instability in 
our numerical calculations.
When we impose the condition, $\rho-3p \ge 0$,
we cannot numerically find cases in which 
$H_{\mbox{\tiny MAX}}$ exceeds Buchdahl's limit, 
as is seen in the 4th and 5th columns in Table \ref{t1}.
Note that the condition, $\beta<-5$, also excludes
all the interesting cases in which
$H_{\mbox{\tiny MAX}}$ exceeds Buchdahl's limit.

Let us find a critical value, $\beta_c<0$, of $\beta$,
such that, for $\beta<\beta_c$,
nonlinear behavior of the scalar field begins to appear.
We numerically calculated $a_s$, $b_s$ and $c_s$
as functions of $n_c$ for $\beta=-4,-5$ and $-6$
under the conditions, $\re'\leq0$ and $\rho-3p\geq0$.
We show $(a_s,b_s)$ and $c_s$ in Fig.\ref{fa} and
Fig.\ref{fi}, respectively.
For $\beta\geq-4$, almost no deviation from general relativity
appears.
In the cases that $\beta=-5$ and $-6$,
these parameters show deviations from general relativity
in which $a_s=b_s$ and $c_s=0$.
Our results are consistent with the previous works \cite{2},\cite{3},
in which $\beta_c$ is found to be $-4.35$.
Note that,
even when $\beta<\beta_c$, our inequality,
$|c_s|\le 2\sqrt{3}/9$, is surely satisfied.
This reconfirms our assertion
that the nonlinear effects are always bounded in this sense.

Now we shall briefly compare our numerical results with those in
 previous works \cite{2},\cite{3}.
The maximum baryonic mass of a star is defined as 
the peak of the $\bar{m}$-$n_c$ relation. 
We numerically found that
the maximum baryonic mass increases from
the general relativistic value, 2.23${\rm M}_{\odot}$, 
to 2.38${\rm M}_{\odot}$ 
and 2.96${\rm M}_{\odot}$ for $\beta=-5$ and $-6$, respectively.
The corresponding radius defined by (\ref{e70}) also increases from 
the general relativistic value, 11.0 km, to 12.0 km 
and 12.9 km for $\beta=-5$ and $-6$, respectively.
In Ref.\cite{2},
a fractional binding energy, $f_{\mbox{\tiny BE}}\equiv
2\bar{m}/b-1$, is used as a measure
of the scalar field contribution to the mass.
We numerically found that the maximum value of $f_{\mbox{\tiny BE}}$
increases from the general relativistic value, 
$f_{\mbox{\tiny BE}}=0.14$, to $f_{\mbox{\tiny BE}}=0.16$ and
$0.22$ for $\beta=-5$ and $-6$, respectively.
Though a slightly different asymptotic value of $\varphi_0$
has been adopted in Ref.\cite{2},
these results are consistent with the previous results. 

The mass-size relation of neutron stars has been
thoroughly studied in general relativity by
solving the Oppenheimer-Volkoff equation,
and it has been found numerically that,
as the equation of state becomes softer,
the mass-to-size ratio becomes larger when its mass is fixed\cite{13}.
In Fig.\ref{fratio1}, we compare, 
under the condition, $\rho-3p\geq0$,
the relations between the mass-to-size ratio and the mass
in general relativity and in scalar-tensor theories.
It is seen in Fig.\ref{fratio1} that
the deviation from general relativity due to the scalar field
begins appearing for $M>1.2{\rm M}_\odot$. 
When $M<1.7{\rm M}_{\odot}$ ($M>1.7{\rm M}_\odot$),
our numerical solutions in the scalar-tensor theory 
correspond to the solutions in general relativity 
with the softer (stiffer) equation of state.
The mass of PSR1913+16 has been evaluated as $1.4{\rm M}_\odot$ [13],
and, if the adopted equation of state is adequate, 
the scalar field contribution to the mass-to-size ratio
is negligibly small when $\beta>-5$.
Further discussion on the equation of state
is left as a future work.

Finally,
we shall derive a redshift formula in the scalar-tensor theories.
A null vector, $k^{\mu}$, 
tangent to the radial null geodesic, and a four-velocity, $U^{\mu}$,
of a static observer are, respectively, given by
\begin{equation}
k^{\mu}= A^{-2}(e^{-\gamma},1,0,0), \ \ \ 
U^{\mu}=(A^{-1}e^{-\gamma /2},0,0,0), \ \ \ 
U^{\mu}U_{\mu} =-1 .
\end{equation}
The frequency, $\omega$, of a light ray is given by
\begin{equation}
\omega=-g_{\mu\nu}k^{\mu}U^{\nu}=A^{-1}e^{-\gamma /2}.
\end{equation}
The redshift, $z$, is then obtained as
\begin{equation}
1+z = \frac{\omega_{\mbox{\tiny source}}}{\omega_{\mbox{\tiny
      observer}}}
=\left. A^{-1}e^{-\gamma /2}\right|_{\mbox{\tiny source}}, 
\end{equation}
where the observer is assumed to be at the spatial infinity, 
$a/\chi \rightarrow 0$. 
By using (3.22),(3.23) and (3.40), we obtain the redshift formula
in the present specific scalar-tensor theory as
\begin{equation}
1+z=(1-a_s)^{-\frac{b_s}{2a_s}}\exp\left\{-\frac{1}{2}\beta
\left[\frac{c_s}{a_s}\ln(1-a_s)\right]^2\right\}=
\frac{1}{b_s}(1-a_s)^{\frac{a_s-2b_s}{2a_s}}
\left(\frac{2M}{R}\right) .
\end{equation}
The maximum value of $z$
depends on $\beta$ and the parameters, $a_s$ and $b_s$, in the
allowed region, $D$, in Fig.1.
Theoretically, the possible maximum value of $z$, 
$z_{\mbox{\tiny max}}$, 
is obtained as $z_{\mbox{\tiny max}}=2$ in general relativity
and $z_{\mbox{\tiny max}}= 164$ and $356$ for $\beta=-5$ and 
$\beta=-6$, respectively.
We compare these values with those obtained in the 
numerical calculations under the condition, $\rho-3p \ge 0$. 
We numerically find that 
$z_{\mbox{\tiny max}}=0.43$ in general relativity, 
and that $z_{\mbox{\tiny max}}=0.44$ and $0.55$ 
for $\beta=-5$ and $-6$, respectively. 
In our numerical solutions with the presently adopted model parameters,
the deviation of the redshift formula from that in general relativity 
remains small compared with the theoretically possible deviation. 
However, the redshift difference, 0.01, numerically found for $\beta=-5$
may be detectable. Therefore the redshift 
measurement of neutron stars will
provide us with a possible tool for the experimental test 
of general relativity. 

\

\section{Summary}

\

We have derived a modified Buchdahl inequality in
scalar-tensor theories of gravity.
As a result, we have obtained two theory-independent
inequalities, $b_s \le 8/9$ and
$|c_s| \le 2\sqrt{3}/9$.
The first inequality corresponds to
the Buchdahl inequality in general relativity.
The second inequality is
characteristic of scalar-tensor theories.
Consequently, even if the scalar field is locally amplified
due to non-perturbative effects in a strong gravitational field,
the characteristic amplitude of the scalar field, $|c_s|$, 
is bounded in this sense.

The modified Buchdahl inequality is then reformulated
to obtain a theory-dependent mass-to-size ratio, $2M/R$,
with an example of the coupling function, $A(\varphi)$,
in a simple form.
If we take $\varphi_0=0$, the mass
in the physical frame is the same as that in general relativity,
$M=b/2$. However, the physical radius, $R$, of the star
can be smaller than the general relativistic one.
As a result, the mass-to-size ratio can exceed not only Buchdahl's
limit but also the black hole limit in contrast to general
relativity.

Our analytic results have been numerically confirmed
when we assume a polytropic equation of state for the matter.
In particular, 
we have found numerical solutions
in which the mass-to-size ratio exceeds both Buchdahl's limit and
the black hole limit. However, 
these theoretically interesting stars could not be 
found numerically under the condition, $\rho-3p \ge 0$,
which is interpreted as a numerical measure of the
onset of the instability of a star. 
Moreover, under this condition, we find numerically that 
any quantitative deviation from general relativity due to the scalar field
remains comparatively small in contrast to our analytic results, where
possible significant effects of the scalar field are expected.
However, as discussed briefly, some measurable effects
in astronomical observations may exist.

Now suppose that a space rocket approaches a massive star
for which $2M/R\gg1$.
If the rocket accidentally goes into
a {\it Schwarzschild radius} of the star defined by $2M$,
a spaceman in the rocket would be resigned to his fate to die.
Now we know that, unfortunately for him,
even if scalar-tensor theories
 describe classical gravity,
he would hardly have a chance
to return alive, because he could hardly meet a real {\it false black hole}.
He has two possible futures, and they are equally tragic:
\begin{itemize}
\item If it is a black hole in terms of general relativity, he can never
  escape.
\item If it is a naked singularity in terms of scalar-tensor theories,
nobody knows what will happen when he touches it.
\end{itemize}

\section*{Acknowledgements}
The authors would like to
thank Dr. K. Oohara for useful discussions regarding numerical
calculations.
They also thank the referee for his careful reading of
the manuscript and valuable comments.

\

\appendix
\section{Observational constraints}

\

In general, the coupling strength, $\alpha(\varphi)$, 
can be an arbitrary function 
of $\varphi$, and in the limit, $\alpha(\varphi) \rightarrow 0$, 
scalar-tensor theories 
approach general relativity.
One defines
\begin{eqnarray}
\alpha_0 & \equiv & \alpha(\varphi_0), 
\label{e17} \\
\beta_0 & \equiv & \left.\frac{d\alpha(\varphi)}
{d\varphi}\right|_{\varphi=\varphi_0},
\label{e18}
\end{eqnarray}
where $\varphi_0$ is the asymptotic value of $\varphi$ 
at spatial infinity.
In the post-Newtonian approximation, the PPN parameters and the effective 
gravitational constant are expressed as follows \cite{8}:
\begin{eqnarray}
1-\gamma_E & = & \frac{2\alpha_0^2}{1+\alpha_0^2}, \label{e19} \\
\beta_E-1 & = & \frac{\beta_0\alpha_0^2}{2(1+\alpha_0^2)^2},
\label{e20} \\
G & = & G_{\ast}A^2(\varphi_0)
\left(1+\alpha^2(\varphi_0)\right). \label{e21}
\end{eqnarray}
General relativity corresponds to the case that 
$\beta_E=\gamma_E=1$ \cite{5},\cite{12}.
Experiments on the time delay and deflection of light in the solar system 
constrain $|1-\gamma_E|$ as \cite{10}
\begin{equation}
|1-\gamma_E| < 2\times 10^{-3},
\label{e22}
\end{equation}
which constrains $\omega(\phi)$ and $\alpha_0$ as
\begin{equation}
\omega > 500 ,  \,\,\,\,\, \alpha_0^2 < 10^{-3}.
\label{e23}
\end{equation}
The lunar-laser-ranging experiments constrain 
$|\beta_E-1|$ as \cite{10}
\begin{equation}
|\beta_E-1| \lesssim 6\times 10^{-4},
\label{e24}
\end{equation}
which only constrains some combination of $\alpha_0$ 
and $\beta_0$. 
Consequently, if $\alpha_0$ tends to zero, 
the constraint on $\beta_0$ is effectively lost.
However, by adopting a specific coupling function, 
$A=\exp\left(\frac{1}{2}\beta\varphi^2\right)$, 
 another constraint on $\beta_0$ 
is obtained
from observations of the binary-pulsars, PSR1913+16, as \cite{3},\cite{10}
\begin{equation}
\beta_0 > -5.
\label{e25}
\end{equation}

When we take $(\ref{cup})$ as a coupling function, 
the coupling strength is $\alpha(\varphi)=\beta\varphi$.
Accordingly, $\alpha_0=\beta\varphi_0$, and
we obtain the constraint on $\varphi_0$.
In this paper we take $\varphi_0 = 0$ for simplicity.

\

\section{An exterior solution}

\

A line element of the Einstein frame in the Just coordinate is \cite{1}
\begin{equation}
ds_{\ast}=-e^{\gamma(\chi)}dt^2+e^{-\gamma(\chi)}d\chi^2+e^{\lambda(\chi)
-\gamma(\chi)}d\Omega^2.
\label{e26}
\end{equation}
The field equations in the exterior space-time are 
\begin{eqnarray}
\gamma''+\gamma'\lambda' & = & 0, \label{e27} \\
-\gamma^{'2}+\gamma'\lambda'-\lambda^{'2}+\gamma''-2\lambda''
& = & 4\varphi^{'2}, \label{fj2} \\
2+e^{\lambda}(\gamma'\lambda'-\lambda^{'2}+\gamma''-\lambda'')
& = & 0, \label{e28} \\
\varphi''+\lambda'\varphi' &= & 0, \label{e29}
\end{eqnarray}
where a prime denotes differentiation with respect to $\chi$.
With (\ref{e27}),(\ref{e28}) and (\ref{e29}), 
the exterior solution can be obtained as
\begin{eqnarray}
e^{\lambda(\chi)} & = & \chi^2-a\chi, \label{e30} \\
e^{\gamma(\chi)} & = & \left(1-\frac{a}{\chi}\right)^{\frac{b}{a}}, 
\label{e31} \\
\varphi(\chi) & = & \varphi_0 + \frac{c}{a}
\ln\left(1-\frac{a}{\chi}\right), \label{e32}
\end{eqnarray}
where $a,b$ and $c$ are constants of integration, and 
$\varphi_0$ denotes the asymptotic value of $\varphi$ at infinity.
With (\ref{fj2}),
one finds
\begin{equation}
a^2-b^2 = 4c^2.
\label{const} 
\end{equation}

The coordinate transformation 
between the Schwarzschild coordinate, $r$, and
the Just coordinate, $\chi$, is given by
\begin{equation}
r^2=\chi^2 \left( 1-\frac{a}{\chi}\right)^{\frac{a-b}{a}}.
\label{e33}
\end{equation}
Note that $r \rightarrow \chi$ at spatial infinity.
In the Schwarzschild coordinate, a line element becomes
\begin{equation}
ds_{\ast}^2=-e^{2\nu(r)}dt^2+e^{2\mu(r)}dr^2
+r^2d\Omega^2.
\label{e34}
\end{equation}
The exterior solution in the Schwarzschild coordinate is given by
\begin{eqnarray}
e^{2\nu(r)} & = & \left(1-\frac{a}{\chi(r)}\right)^{\frac{b}{a}},
\label{e35} \\
e^{2\mu(r)} & = & \left(1-\frac{a}{\chi(r)}\right)
\left(1-\frac{a+b}{2\chi(r)}\right)^{-2}.
\label{e36}
\end{eqnarray}
Asymptotic behavior of the exterior solution 
at spatial infinity are as follows:
\begin{eqnarray}
e^{2\nu(r)} & \longrightarrow & 1-\frac{b}{r}, \label{e37} \\
e^{2\mu(r)} & \longrightarrow & 1+\frac{b}{r}, \label{e38} \\
\varphi(r) & \longrightarrow & \varphi_0 - \frac{c}{r}. \label{e39}
\end{eqnarray}

\

\section{An interior solution: Numerical methods}

\

Using the variables, $m_\ast(r)$ and $\nu(r)$, defined by 
\begin{equation}
f_\ast(r)\equiv e^{2\nu(r)},~~~~~
h_\ast(r)\equiv\left[1-\frac{2m_\ast(r)}{r}\right]^{-1},
\label{e44}
\end{equation}
the field equations $(\ref{f1})\sim(\ref{f4})$ become
\begin{eqnarray}
\frac{dm_{\ast}}{dr} & = & 4\pi G_{\ast}A^4(\varphi)r^2\rho
+\frac{1}{2}r(r-2m_{\ast})\psi^2, \label{e45} \\
\frac{d\nu}{dr} & = & \frac{m_{\ast}+4\pi G_{\ast}A^4(\varphi)r^3p}
{r(r-2m_{\ast})}+\frac{1}{2}r\psi^2\equiv\Phi(r), \label{e46} \\
\frac{d\varphi}{dr} & = & \psi, \label{e47} \\
\frac{d\psi}{dr} & = & \frac{4\pi G_{\ast}A^4(\varphi)r}
{r-2m_{\ast}}\left[\alpha(\varphi)(\rho-3p)+(\rho-p)r\psi\right]
-\frac{2(r-m_{\ast})}{r(r-2m_{\ast})}\psi, \label{e48} \\
\frac{dp}{dr} & = & -(\rho+p)\left(\Phi
+\alpha(\varphi)\psi\right). \label{e49}
\end{eqnarray}
The total baryon mass measured in the physical frame is
\begin{equation}
\bar{m}=
m_b\int n\sqrt{-g}u^0 d^3x
=
m_b\int^{r_s}_{0}4\pi nA^3(\varphi)r^2
\left(1-\frac{2m_{\ast}}{r}\right)^{-\frac{1}{2}}dr.
\label{e50}
\end{equation}
Given the equation of state,
we can numerically integrate the above field equations outward
from the center, $r=0$, with the boundary conditions as follows:
\begin{equation}
\left.
\begin{array}{ccc}
m_{\ast}(0) & = & 0, \\
\varphi(0)  & = & \varphi_c, \\
\psi(0)     & = & 0, \\
p(0)        & = & p_c, \\
\rho(0)     & = & \rho_c, \\
\end{array}
\right\}
\label{e53}
\end{equation}
where $\rho_c$ and $p_c$ are given by replacing
$n$ in $(\ref{eos1})$ and $(\ref{eos2})$
with $n_c\equiv n(0)$.
The surface of a star, $r=r_s$, is determined by
the condition, $p(r_s)=0$.
A numerically obtained interior solution is to be
matched to the exterior one by the conditions \cite{2},\cite{3}:
\begin{eqnarray}
\varphi_0  & = & \varphi_s + \frac{\psi_s}{\sqrt{{{\nu'}_s}^2+{\psi_s}^2}}
\tanh^{-1}\left(\frac{\sqrt{{{\nu'}_s}^2+{\psi_s}^2}}{{{\nu'}_s}+1/r_s}
\right), \label{e55} \\ 
b & = & 2{r_s}^2{\nu'}_s\sqrt{1-\frac{2m_{\ast s}}{r_s}} \,
\exp\left(-\frac{\nu'_s}{\sqrt{{{\nu'}_s}^2+{\psi_s}^2}}
\tanh^{-1}\left(\frac{\sqrt{{{\nu'}_s}^2+{\psi_s}^2}}{{{\nu'}_s}+1/r_s}
\right)\right), \label{e56} \\
c & = & \frac{\psi_s}{2{\nu'}_s}b, \label{z1} \\
a & = & \sqrt{b^2+4c^2}, \label{z2}
\end{eqnarray}
where a prime denotes differentiation with respect to $r$, and 
the subscript, $s$, refers to quantities evaluated at the surface, $r_s$.
The central value of $\varphi$, $\varphi_c$, is
chosen such that we have $\varphi_0=0$.
 
\



\newpage

\begin{figure}
\begin{center}
\hspace*{0.5cm}
\caption{Each term of (\ref{abc}) in $\re'(r)$ is shown
for $\beta=-5$ and $n_c/n_0 = 10$.
Horizontal and vertical axes denote, respectively,
the radial coordinate, $r$, in the unit of $10\mbox{km}$,
and the 1st, 2nd and 3rd terms in $(\ref{abc})$.}
\label{f-5.2}
\end{center}
\end{figure}

\begin{figure}
\begin{center}
\hspace*{0.5cm}
\caption{We compare $\phi(r)$, $\rho(r)$ and $\varphi(r)$
  in the case that $\beta=-5$ and $n_c/n_0=7.9$.
Horizontal and vertical axes denote, respectively,
the radial coordinate, $r$, in the unit of $10\mbox{km}$,
and $\rho(r)/(m_bn_0)$, $\phi(r)$ and  $\varphi(r)$.
  On a thin dotted line, $G\equiv1/\phi_0=1$, i.e., general
  relativity.}
\label{f-5.4}
\end{center}
\end{figure}

\begin{figure}
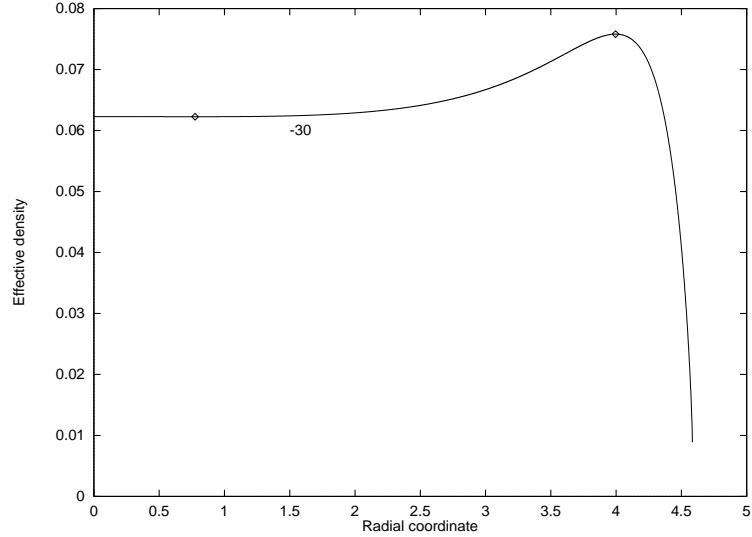

\begin{center}
\hspace*{0.5cm}
\caption{The effective density, $\re(r)$, is shown
in the case that $\beta=-30$ and $n_c/n_0=10$.
Horizontal and vertical axes denote, respectively,
the radial coordinate, $r$, in the unit of $10\mbox{km}$, and
the effective density, $\re(r)/(m_bn_0)$.
It is seen that the assumption, $\red(r) \le 0$, is partially
violated between two rectangles.}
\label{fb}
\end{center}
\end{figure}

\begin{figure}
\begin{center}
\hspace*{0.5cm}
\caption{Each term of (\ref{abc}) in $\re'(r)$ is shown
for $\beta=-30$ and $n_c/n_0 = 10$.
Horizontal and vertical axes denote, respectively,
the radial coordinate, $r$, in the unit of $10\mbox{km}$,
and the 1st, 2nd and 3rd terms in $(\ref{abc})$.}
\label{fdif}
\end{center}
\end{figure}

\begin{figure}
\begin{center}
\hspace*{0.5cm}
\caption{We compare $\phi(r)$, $\rho(r)$ and $\varphi(r)$
  in the case that $\beta=-30$ and $n_c/n_0=10$.
Horizontal and vertical axes denote, respectively,
the radial coordinate, $r$, in the unit of $10\mbox{km}$,
and $\rho(r)/(m_bn_0)$, $\phi(r)$ and  $\varphi(r)$.
  On a thin dotted line, $G\equiv1/\phi_0=1$, i.e., general
  relativity.}
\label{f-30.cl}
\end{center}
\end{figure}

\begin{figure}
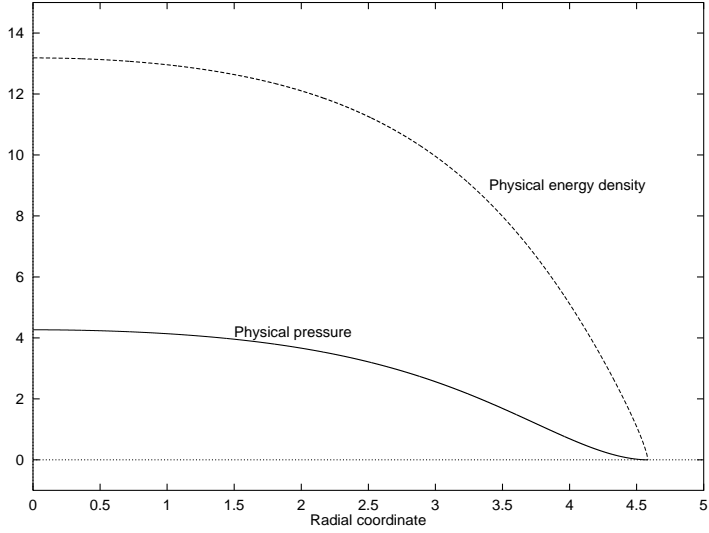

\begin{center}
\hspace*{0.5cm}
\caption{We show $\rho(r)$ and $p(r)$ in the physical frame
for $\beta=-30$ and $n_c/n_0 = 10$.
Horizontal and vertical axes denote, respectively,
the radial coordinate, $r$, in the unit of $10\mbox{km}$,
and $\rho(r)/(m_bn_0)$ and $p(r)/(m_bn_0)$.
Though the assumption, $\red(r)\leq0$, is violated in the Einstein
frame,
the conditions, $\rho(r) - 3p(r) \ge 0$, $\rho'(r)\leq0$ and
$p'(r)\leq0$ are all satisfied in the physical frame.}
\label{fc}
\end{center}
\end{figure}

\begin{figure}
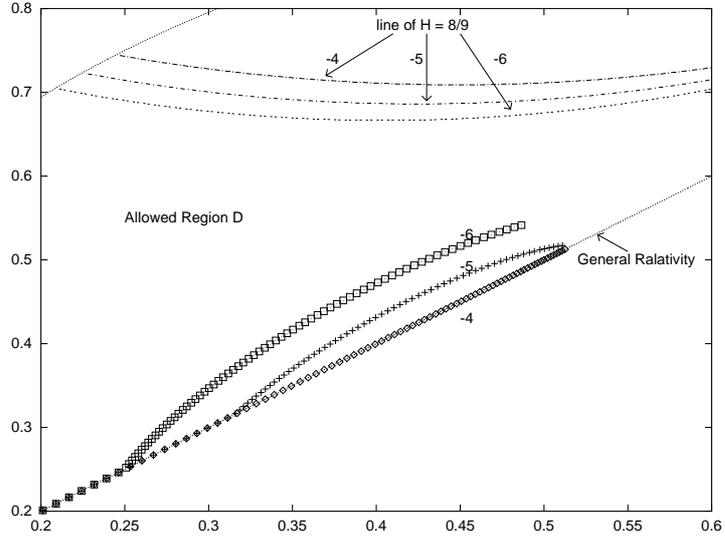

\begin{center}
\hspace*{0.5cm}
\caption{We show the parameters, $(a_s,b_s)$, in each equilibrium
  solution
  for $n_c/n_0=2.5\sim10.3$.
We take $\beta=-4,-5$ and $-6$, and impose the conditions,
$\rho(r)-3p(r)\geq0$ and $\re'(r)\leq0$.
Horizontal and vertical axes denote, respectively, $b_s$ and $a_s$. }
\label{fa}
\end{center}
\end{figure}

\begin{figure}
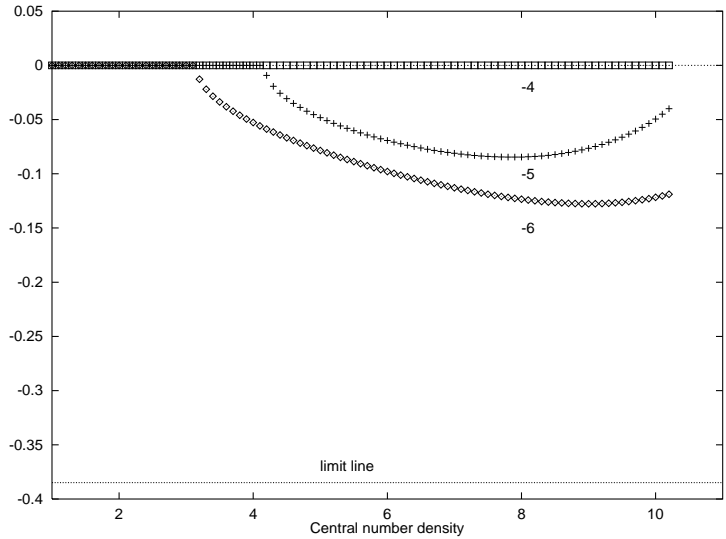

\begin{center}
\hspace*{0.5cm}
\caption{We show the parameter, $c_s$, in each equilibrium solution
  for $n_c/n_0=1.0\sim10.3$.
We take $\beta=-4,-5$ and $-6$, and impose the conditions,
$\rho(r)-3p(r)\geq0$ and $\re'(r)\leq0$.
Horizontal and vertical axes denote, respectively,
$n_c/n_0$ and $c_s$. A horizontal thin dotted line denotes
the limit on $c_s$, i.e., $c_s=-2\sqrt{3}/9$.
}
\label{fi}
\end{center}
\end{figure}

\begin{figure}
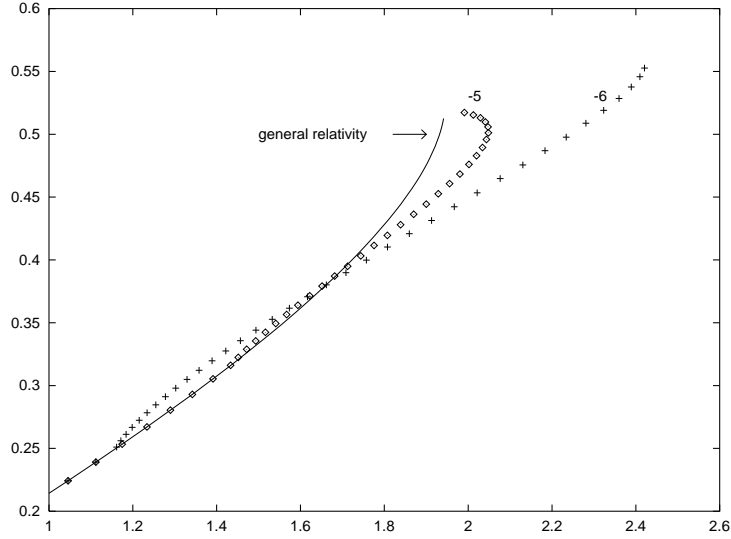

\begin{center}
\hspace*{0.5cm}
\caption{We show the relation between the mass-to-size ratio and the mass.
We take $\beta=-5$ and $-6$, and impose the conditions,
$\rho(r)-3p(r)\geq0$ and $\re'(r)\leq0$.
Horizontal and vertical axes denote, respectively,
$M/{\rm M}_{\odot}$, and $2M/R$. 
The solid line represents the mass-to-size ratio 
in general relativity.  
}
\label{fratio1}
\end{center}
\end{figure}

\begin{table}[htbp]
\begin{center}
\begin{tabular}{|ccccc|}
\hline
$\beta$ & $n_c/n_0$ & $\red\le0$ & $\rho-3p\ge0$ & $H_{\mbox{\tiny
    max}}$  \\
\hline
\hline
$-12.07$ & $0.1\sim10.3$ & $\bigcirc$ & $\bigcirc$ & $0.679$ \\
         & $10.3\sim11.2$ & $\bigcirc$ & $\times$   & $1.018$ \\
$-11.0$  & $0.1\sim10.3$ & $\bigcirc$ & $\bigcirc$ & $0.666$ \\
         & $10.3\sim11.3$ & $\bigcirc$ & $\times$   & $0.919$ \\
$-10.0$  & $0.1\sim10.3$ & $\bigcirc$ & $\bigcirc$ & $0.651$ \\
         & $10.3\sim11.4$ & $\bigcirc$ & $\times$   & $0.834$ \\
$-6.0$   & $0.1\sim10.3$ & $\bigcirc$ & $\bigcirc$ & $0.556$ \\
         & $10.3\sim11.2$ & $\bigcirc$ & $\times$   & $0.569$ \\
$-5.0$   & $0.1\sim10.3$ & $\bigcirc$ & $\bigcirc$ & $0.517$ \\
         & $10.3\sim15.7$ & $\bigcirc$ & $\times$   & $0.567$ \\
$-4.0$   & $0.1\sim10.3$ & $\bigcirc$ & $\bigcirc$ & $0.514$ \\
         & $10.3\sim14.9$ & $\bigcirc$ & $\times$   & $0.562$ \\
$ 0.0$   & $0.1\sim10.3$ & $\bigcirc$ & $\bigcirc$ & $0.514$ \\
         & $10.3\sim14.9$ & $\bigcirc$ & $\times$   & $0.562$ \\
\hline
\end{tabular}
\end{center}
\caption{
We summarize our numerical results. In the 1st column,
$\beta$ is given. In the 2nd column, we give a range of
$n_c$ from our numerical studies. In the 3rd and 4th
columns, we indicate, respectively, whether the assumption,
$\red\leq0$, and the condition, $\rho-3p\geq0$, are satisfied.
In the 5th column, the maximum mass-to-size ratio
is shown for each $\beta$.
}
\label{t1}
\end{table}


\begin{thebibliography}{99}
\bibitem{4}C. Brans and R. H. Dicke,
         Phys. Rev. $\mathbf{124}$, 925 (1962).
\bibitem{15}P. G. Bergmann, Int. J. Theor. Phys. $\mathbf{1}$, 25
  (1968).
\bibitem{14}R. V. Wagoner, Phys. Rev. $\mathbf{D1}$, 3209, (1970).
\bibitem{5}C. M. Will,
         {\it Theory and Experiment in Gravitational Physics},
         (Cambridge University Press, Cambridge, 1993).
\bibitem{16}M. B. Green, J. H. Schwartz and E. Witten,
            {\it Superstring Theory vols. 1,2},
  (Cambridge University Press, Cambridge, 1987).
\bibitem{2}T. Damour and G. Esposito-Far\`{e}se,
         Phys. Rev. Lett. $\mathbf{70}$, 2220 (1993).
\bibitem{3}T. Damour and G. Esposito-Far\`{e}se,~ {\it preprint}
         gr-qc/9602056.
\bibitem{11}H. A. Buchdahl, Phys. Rev. $\mathbf{116}$, 1027 (1959).
\bibitem{7}S. Weinberg,
         {\it Gravitation and Cosmology},
         (Wiley, New York, 1972).
\bibitem{6}R. M. Wald,
         {\it General Relativity},
         (The University of Chicago Press, Chicago and London, 1984).
\bibitem{8}T. Chiba, T. Harada and K. Nakao,
  Prog. Theor. Phys. Suppl. $\mathbf{128}$, 335 (1997).
\bibitem{20}D. L. Lee, Phys. Rev. $\mathbf{D10}$, 2374 (1974).
\bibitem{13}S. L. Shapiro and S. A. Teukolsky,
  {\it Black Holes, White Dwarfs and Neutron Stars}, (Wiley, 1983).
\bibitem{10}G. Esposito-Far\`{e}se, {\it preprint}~ gr-qc/9612039.
\bibitem{12}C. W. Misner, K. S. Thorne and J. A. Wheeler,
               {\it Gravitation}, (Freeman, San Francisco, 1973).
\bibitem{1}T. Damour and G. Esposito-Far\`{e}se,
         Class. Quant. Grav. $\mathbf{9}$, 2093 (1992).
\end{thebibliography}
\end{document}